\documentstyle[11pt,IAUS212,twoside,epsf]{article}

\begin{document}

\title{The Evolutionary Stage of 5 Southern Galactic Unclassified B[e] Stars}
\author{Marcelo Borges Fernandes, Francisco X. de Ara\'ujo}
\affil{Observat\'orio Nacional, Rua General Jos\'e Cristino, 77, CEP: 20921-400, 
S\~ao Crist\'ov\~ao, Rio de Janeiro, Brasil}
\author{H.J.G.L.M. Lamers}
\affil{Astronomical Institute, Utrecht University, P.O. Box 80000, NL-3508, TA, 
Utrecht, The Netherlands}

\begin{abstract}
The spectra of stars with the B[e] phenomenon are dominated by features that are 
related to physical conditions of circumstellar material around these objects 
and are not intrinsic to the stars. Because of this, they form a very 
heterogeneous group. This group contains objects with different evolutionary 
stages. Lamers et al. (1998) have suggested a new designation with five 
sub-groups, which indicate the evolutionary stage. They are: supergiants, 
pre-main sequence or Herbig Ae/Be, compact planetary nebulae, symbiotic and 
unclassified. The unclassified group has many objects that need a better study 
to resolve their evolutionary status. Forbidden lines can be a useful tool to 
solve this problem. They can give 
informations about chemical composition, ionization and density of the 
circunstellar medium and probably the evolutionary phase of these objects. We 
analize spectra of some galactic objects, obtained with FEROS and B\&C 
spectrograph at 1.52 telescope in ESO (La Silla-Chile), with a 
special focus on the forbidden lines. We have studied the spectra of 5 B[e] 
stars of uncertain evolutionary stage. We find that one of them is a pre-WN 
star, the other four are supergiant B[e] stars.
\end{abstract}

\section{Forbidden Lines}

The presence of some forbidden lines is a criterion to distinguish differents 
groups of massive stars. For 
example, LBV and sgB[e] have similar spectral characteristics, however, [OI] 
lines are present only in the sgB[e]  
(Zickgraf, 1989). Below, we have a table showing the presence (Yes) or not (No) 
of some important forbidden lines. Figure 1 shows the profiles of forbidden and 
permitted NII lines in the spectra of HD 326823.  

\vskip 0.4truecm
\begin{center}
{\small
\begin{tabular}{|c|c|c|c|c|c|c|} \hline
\multicolumn{1}{|c|}{ Objects } &
\multicolumn{1}{|c|}{ [OI] } &
\multicolumn{1}{|c|}{ [OII] } & 
\multicolumn{1}{|c|}{ [OIII] } &
\multicolumn{1}{|c|}{ [SII] } &
\multicolumn{1}{|c|}{ [NII] } &
\multicolumn{1}{|c|}{ [FeII] } \\
\hline \hline
HD 87643 & Yes & No & No & Yes & No & Yes\\
Hen 3-847 & Yes & No & No & Yes & Yes & Yes \\
GG Car & Yes & No &  No & No & Yes & Yes \\
MWC 300 & Yes & No & No  & Yes  & Yes & Yes \\ 
HD 326823 &  No & No & No & Yes & Yes & Yes \\ \cline{1-3} \hline
\end{tabular}
}
\end{center}

\vskip 60truecm
\vbox{\includegraphics{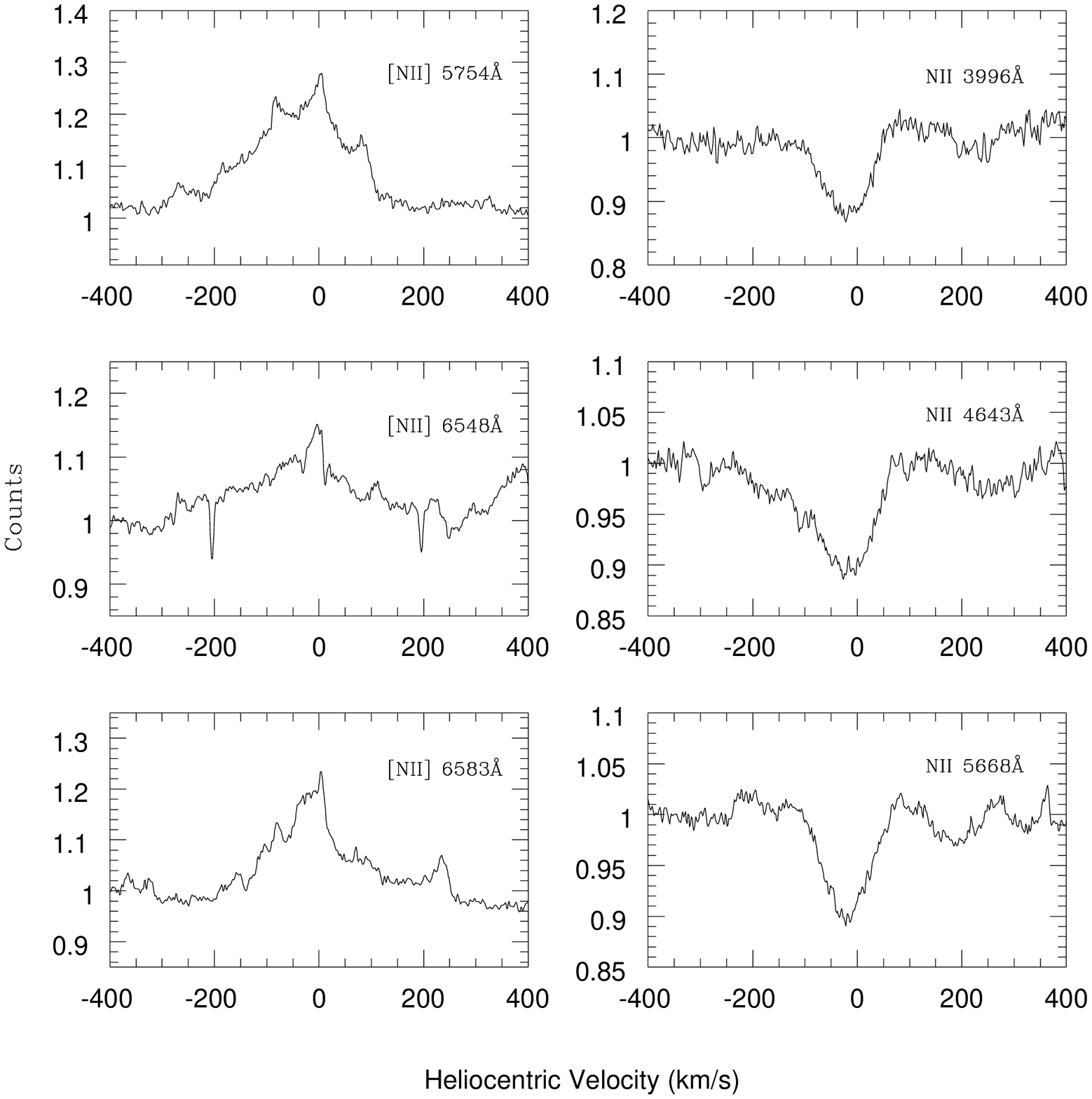}}
\vskip 5.5truecm
\noindent {\footnotesize {\bf \it Figure 1 -}Profiles of NII lines in the high 
resolution spectrum (FEROS-agreement ESO/ON) of HD 326823. The permitted 
lines are in absorption, indicating a 
photospheric origin, and the forbidden are in emission, indicating a 
circumstellar origin. }

\section{The Evolutionary Phase of unclB[e] Stars}

From the comparison with spectra of well-stabilished groups, it is possible to 
derive the evolutive stage of these 5 unclB[e].

{\bf \underline {HD 326823}} - The He 
overabundance indicates that it is 
an evolved star. In agreement with the literature, we suggest that is a pre-WN 
star with the B[e] phenomenon (Fernandes et al. 2001). This suggests a new 
class of B[e] stars.

{\bf \underline {HD87643}} - There is a doubt if 
it is a pre-main 
sequence or a supergiant object. The absence of inverse P-Cygni profiles and 
absence of a strong 
variability suggest an evolved star, as sgB[e].

{\bf \underline {Hen 3-847}} - This object used 
to 
be classified as a pre-main sequence star. However, our spectra do not show any 
characteristic feature of young objects. We suggest that it is a sgB[e] 
candidate.

{\bf \underline {GG Car}} - This star was classified 
as sgB[e], but some authors consider it as a binary system, with 
a B star and a K star. Our spectra do not show any 
feature from the late type star. So we suggest that it is a sgB[e].

{\bf \underline {MWC 300}} - This object is classified, 
in the literature, as sgB[e] candidate. Our analysis of its spectrum 
confirms that suggestion.

We have already started a study of the physical properties and parameters of the 
circumstellar material of 
those objects. The full result of this study will be published in \aap \ by the 
same authors.

\end{document}